# Design and Modeling of a Simple-Structured Continuously Variable Transmission Utilizing Shape Memory Alloy Superelasticity for Twisted String Actuator[†]

Chanchan Xu[a], Shuai Dong[b*] and Xiaojie Wang[c*]

Twisted String Actuators (TSAs) are widely used in robotics but suffer from a limited range of Transmission Ratio (TR) variation, restricting their efficiency under varying loads. To overcome this, we propose a novel lightweight, simple-structured Continuously Variable Transmission (CVT) mechanism for TSA utilizing Shape Memory Alloy (SMA) superelasticity. The CVT mechanism consists solely of a pair of highly lightweight superelastic SMA rods connecting the ends of twisted strings. These rods deform under external loads, adjusting the inter-string distance to enable continuous TR variation. We develop a comprehensive theoretical model that integrates three critical nonlinearities: SMA material nonlinearity (variable elastic modulus), geometric nonlinearity (large deflections), and axial deformation dynamics. This model combines fundamental TSA kinematics with SMA rod deformation, forming a boundary-value problem governed by highly nonlinear implicit differential equations. A numerical solution approach, incorporating the shooting method and Newton–Raphson iteration method and Runge-Kutta method, is used to solve these equations. Numerical simulations and experimental validation demonstrate the proposed CVT's ability to dynamically adapt its TR to varying loads, achieving both high-speed motion under light loads and substantial output force under heavy loads. Experimental results confirm the model's reliability, and parameter studies using this model provide key insights for performance optimization. This innovation offers a promising solution for diverse robotic applications, highlighting the mechanism's adaptability.

## 1 Introduction

TSA is a prominent class of soft actuators that efficiently convert a motor's rotational motion into linear motion by twisting one or more strings [1,2]. A typical TSA system comprises an electric motor, a load, and one or more strings that function as a transmission mechanism, linking the motor to the load. TSA is renowned for their simple yet compact design, lightweight construction, high transmission efficiency, substantial output force, energy efficiency, inherent compliance, cost-effectiveness, and long-distance power transmission capability [3,4]. Owing to these advantages, they have been widely adopted in diverse applications, including robotic hands [5,6], robot exoskeletons [7], tensegrity robots [8], and soft robotic systems [9,10].

However, despite their benefits, TSA encounters a significant challenge in practical applications due to a limited range of TR(the ratio of the motor speed to the linear velocity of the driven load, or the ratio of the driven load force to the motor output torque) variation [11,12]. Ideally, robotic actuators should offer both high-speed motion under no-load and substantial supporting force under loaded. While using excessively powerful actuators can meet both speed and force requirements, it often results in energy inefficiency. A CVT mechanism can overcome this challenge by enabling robotic systems to achieve both high-speed motion with small load and substantial supporting force for large load, thereby maximizing actuator efficiency [13,14]. To expand the range of TR variation in TSA, a CVT mechanism for TSA is essential to tailor the TR according to operational loads.

Researchers have developed various CVT mechanisms integrated with TSA, primarily by modifying string radius, input torque, or inter-string distance. The variable string radius approach is widely adopted, encompassing dual-mode mechanisms [12,15], load-adaptive hoisting mechanisms [16], and active-type CVT [11]. Meanwhile, mechanisms modulating string input torque include elastomeric CVT [17] and adaptable toroidal CVT [18]. To simplify CVT for TSA, a simple-structured design can adjust TR by using springs to vary the distance between two strings based on external loads [19]. Despite advances in CVT for TSA, existing solutions remain hindered by various combinations of excessive weight, bulky designs, structural complexity, discontinuous TR adjustment, lack of compliance, and energy inefficiency.

To address this need, we have developed a lightweight, simple-structured, low-cost, and compliant CVT for TSA [20]. This CVT mechanism consists solely of a pair of highly lightweight SMA hyperelastic slender rods connected to both ends of the twisted strings. It can deform in response to external load, thereby adjusting the distance between the two twisted strings to enable the TR to continuously adapt to varying load conditions. Through comparative analysis with existing TSA-based CVTs, our proposed CVT for TSA emerges as the lightest, simplest in structure, and most cost-effective option. Moreover, by leveraging SMA's hy-perelastic properties, our CVT achieves bidirectional adjustability, high energy efficiency, and enhanced durability while maintaining inherent compliance—providing critical protection to both motor and load against operational impacts and collisions.

[a] *School of Artificial Intelligence, Anhui Polytechnic University, Wuhu, China; E-mail: ccxu@ahpu.edu.cn*
[b] *Institute of Humanoid Robots, Department of Precision Machinery and Precision Instrumentation, University of Science and Technology of China, Hefei, China; E-mail: shuaizhi@ustc.edu.cn*
[c] *Institute of Intelligent Machines, Hefei Institutes of Physical Science, Chinese Academy of Sciences, Hefei, China; E-mail: jwang@iamt.ac.cn*

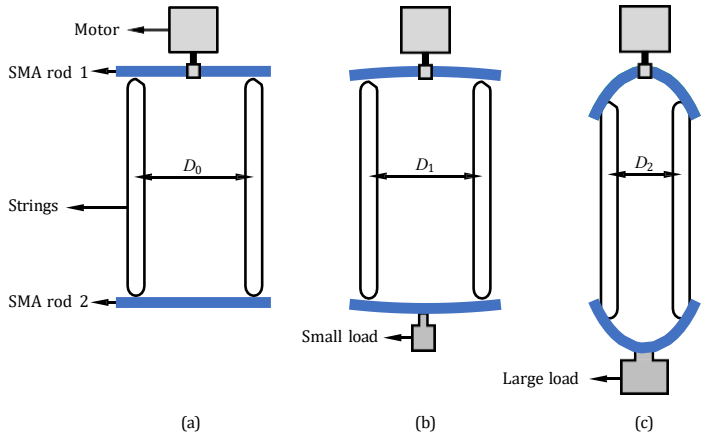

Fig. 1 The proposed TSA-based CVT mechanism. (a) The state without load. (b) The state with a small load. (c) The state with a large load.

Previous studies by our group modeled this mechanism, assuming constant elastic modulus for the SMA material to simplify calculations. While this simplification yields reasonably accurate predictions under light loads, it proves poor prediction accuracy for larger load conditions. This limitation arises from SMA's nonlinear stress-strain behavior—characterized by a distinctive plateau phase where stress remains constant despite strain variation—resulting in inherently variable elastic modulus [21,22]. To develop a comprehensive theoretical model for the proposed TSA-based CVT mechanism, this work simultaneously incorporates three critical nonlinearities:

1. Material nonlinearity (variable modulus of elasticity of SMA),
2. Geometric nonlinearity (large-deformation effects),
3. Axial deformation dynamics of SMA rods.

This model integrates the fundamental TSA kinematic model with the deformation model of SMA rods, thereby forming a complex boundary-value problem that is governed by highly nonlinear implicit differential equations.

The paper is structured as follows: Section II elaborates on the proposed design concept and working principle. Section III delves into establishing the theoretical model of the proposed TSA-based CVT mechanism, including the TSA fundamental model and the superelastic SMA rod model. Numerical simulation results and discussions are presented in Section IV. Experimental analysis is conducted in Section V, followed by the conclusion in Section VI.

## 2 Design concept and working principle

The structure and working principle of the proposed TSA-based CVT mechanism are depicted in Figure 1. This mechanism comprises a motor, a pair of superelastic SMA rods (SMA rod 1 and SMA rod 2), two strings, and a load. SMA rod 1 is connected to the motor's output shaft, while SMA rod 2 is connected to load parallel to SMA rod 1. Both SMA rods are linked by two strings parallel to the motor's output shaft.

When the motor drives SMA Rod 1 to rotate, the two strings are driven by SMA Rod 1 to wind around each other, shortening their length and causing SMA Rod 2 and the load to move linearly. Additionally, variations in the distance between the strings have an impact on the contraction rate of the strings, and there is a positive correlation between these two parameters. Hence, when the load on SMA rod 2 is lighter, deformation is nearly absent, allowing for a considerable distance between the strings. This facilitates rapid contraction, resulting in a small TR. Conversely, with heavier load, SMA rod 2 undergoes bending deformation, narrowing the string distance and slowing the contraction rate, thereby increasing the TR and augmenting load capacity. Additionally, as the load magnitude rises, the bending deformation intensifies, further reducing the string distance and enhancing both the TR and load capacity. Consequently, the proposed TSA-based CVT mechanism adaptively adjusts its TR in response to load weight, ensuring optimal driving speed and load capacity of the TSA under different load conditions.

## 3 Theoretical analysis

In this section, we construct a mathematical model for the proposed TSA-based CVT mechanism, elucidating the connection between the mechanism's input and output displacements to characterize the TR of the entire mechanism. This model amalgamates the TSA fundamental model with the deformation model of superelastic SMA rod. We will provide a comprehensive account of the individual establishment and integration processes of these two models.

### 3.1 TSA fundamental model

The operation process of the TSA-based CVT mechanism is illustrated in Figure 2. To avoid an abrupt contraction of strings leading to an uncomfortable linear motion of load, the string is pre-twisted half a turn, as illustrated in Figure 2(a). Once this pre-twisting is applied, we can calculate the initial distance, denoted as $X_0$, which represents the vertical distance between the connection points of the two SMA rods and the string on the same side, using the following formula:

$$X_0 = \sqrt{L_0^2 - D_0^2} \qquad (1)$$

where, $L_0$ represents the initial length of the string without load force, and $D_0$ is the initial horizontal distance between the connection points of the twisted strings and both ends of the SMA rod without load force.

When a workload weighting $F_z$ is applied to SMA rod 2, as depicted in Figure 2(b), the strings undergo tension, causing them to elongate from $L_0$ to $L_1$. Additionally, this tension induces bending forces on both SMA rod 1 and SMA rod 2, resulting in downward and upward bending deformations, respectively. This transformation causes point $A_0$, where the string connects to SMA rod 1, to shift to $A_1$, leading to a decrease in the horizontal distance between the ends of the SMA rod and the string connection from $D_0$ to $D_1$, and a vertical movement of the ends of SMA rod 1 and the string connection downward by $y_1$. Given the symmetry between the upward bending deformation of SMA rod 2 and the downward bending of SMA rod 1, the ends of SMA rod 2 and the string connection also experience an identical but upward displacement of $y_1$. Under the combined effect of the strings and SMA rods, the vertical distance between the connection points of the two SMA rods and the strings on the same side shifts from $X_0$ to $X_1$.

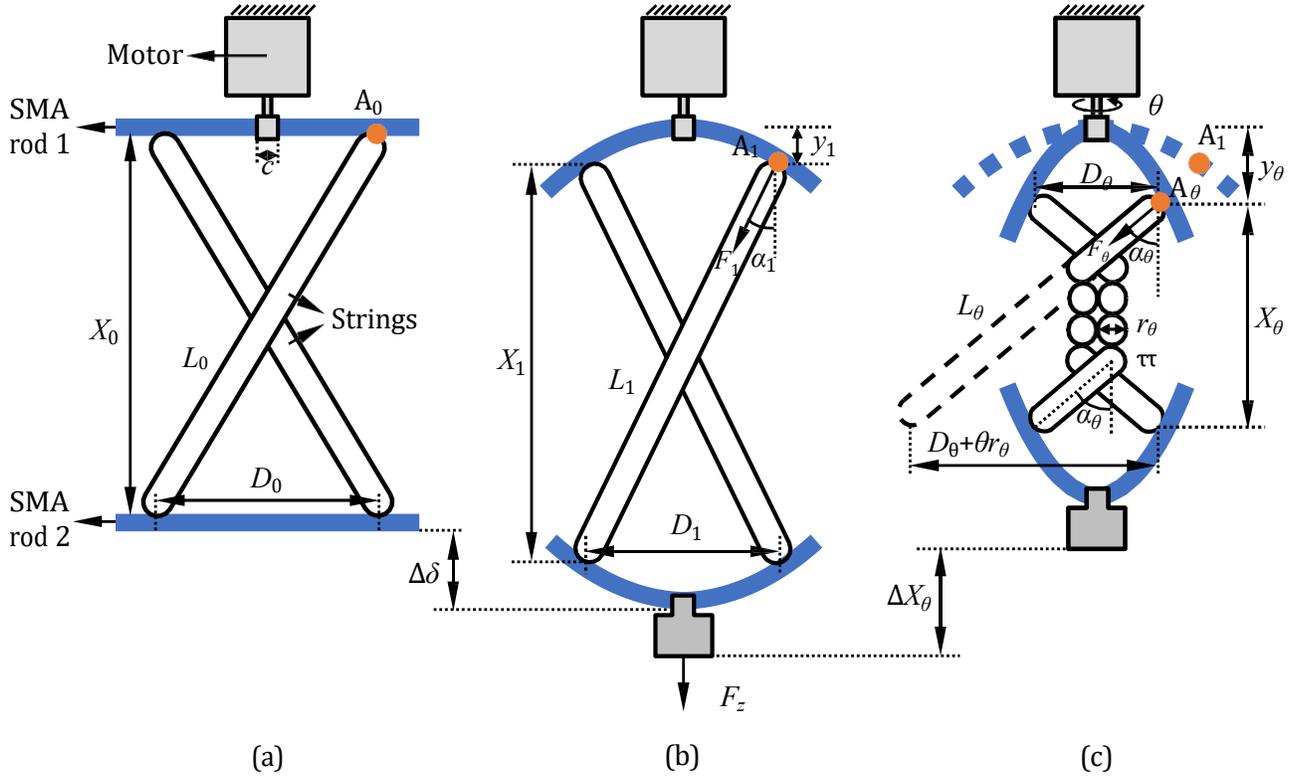

Fig. 2 The proposed TSA-based CVT mechanism with pre-twisted half a turn for initial state without load (a) and with load (b). (c) The state with load after the motor makes a certain number of turns $\vartheta$.

Therefore, after attaching a workload to SMA rod 2, the workload generates a downward displacement given by:

$$\Delta \delta = X_1 - X_0 + 2y_1 \qquad (2)$$

As the motor rotates by an angle $\vartheta$, the strings undergo a corresponding twisting motion, taking on a cylindrical shape with a radius of $r_\vartheta$ in their midsection, as shown in Figure 2(c). This leads to an increase in tension within the strings and further bending of the SMA rods, causing $A_1$, $L_1$, $D_1$, and $y_1$ to transform into $A_\vartheta$, $L_\vartheta$, $D_\vartheta$, and $X_\vartheta$, respectively. Therefore, when the motor rotates from half a turn to angle $\vartheta$, the resulting displacement of the load ($\Delta X_\vartheta$) can be expressed as

$$\Delta X_\vartheta = X_1 - X_\vartheta - 2(y_\vartheta - y_1) \qquad (3)$$

Now suppose one of the strings is unwound geometry of a cylinder, as shown in Figure 2(c), so $X_\vartheta$ can be calculated by

$$X_\vartheta = L_\vartheta \cos \alpha_\vartheta \qquad (4)$$

where $\alpha_\vartheta$ is the angle between the strings and the motor axis direction when it is twisted by an angle $\vartheta$. We assume that the helix angle of the formed cylinder is evenly distributed across the twists of the strings, thus, the helix angle equals the angle between the strings and the motor axis direction $\alpha_\vartheta$. So

$$\cos \alpha_\vartheta = \frac{X_\vartheta}{\sqrt{X_\vartheta^2 + (D_\vartheta + \vartheta r_\vartheta)^2}} \qquad (5)$$

The $r_\vartheta$ is variable proved in [23]. However, considering the strings' elasticity and their deformation under tension, we accounted for the applied load force's impact on the radius. As a result, we approximate the radius $r_\vartheta$ of the cylindrical shape formed by the middle section of the strings using the following relationship:

$$r_\vartheta = \left( \frac{1}{k_1 F_z} + k_2 + k_3 \sqrt{X_0/X_\vartheta} \right) r_0 \qquad (6)$$

where $k_1$, $k_2$, and $k_3$ denote approximation coefficients that depend on the structure and mechanical properties of the strings, and $r_0$ represents the diameter of the strings in their unloaded state.

In addition, $L_\vartheta$ can be obtained by

$$L_\vartheta = L_0 + \frac{F_\vartheta}{K} \qquad (7)$$

$$F_\vartheta = \frac{F_z}{2 \cos \alpha_\vartheta} \qquad (8)$$

Here, $F_\vartheta$ denotes the axial force in the string under a twist angle $\vartheta$, while $K$ stands for the string's stiffness coefficient.

Therefor, equation (4) turns to:

$$X_\vartheta = L_0 \cos\alpha_\vartheta + \frac{F_z}{2K} \quad (9)$$

The above describes the establishment process of the TSA fundamental model. However, solving for $\Delta X_\vartheta$ depends on known values of $D_1, y_1, D_\vartheta$, and $y_\vartheta$. Since $D_1$ and $y_1$ represent the values of $D_\vartheta$ and $y_\vartheta$ when $\vartheta = \pi$, we only need to obtain $D_\vartheta$ and $y_\vartheta$ for any $\vartheta$. As $D_\vartheta$ and $y_\vartheta$ are parameters reflecting the bending deformation of SMA rods, we will conduct the bending deformation analysis on the SMA rods in the following section and combine it with the aforementioned TSA fundamental model to solve for $\Delta X_\vartheta$.

## 3.2 Superelastic SMA rod model
### 3.2.1 Basic analysis

We consider the entire process of motor rotation in the TSA-based CVT mechanism to be quasi-static, implying that SMA rod 1 and rod 2 maintain force equilibrium throughout the process. Due to the symmetrical loading deformation of SMA rod 1 and rod 2, only SMA rod 1 is analyzed. Therefore, after the motor completes a certain number of rotations $\vartheta$, the deformation and force equilibrium state of SMA rod 1 in Figure 2(c) alone is as shown in Figure 3.

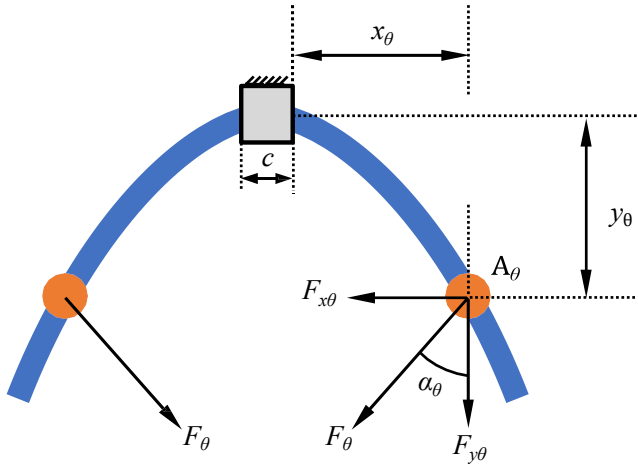

Fig. 3 The bending deformation analysis of SMA rod 1 in Figure 2(c) alone

SMA rod 1 is fixed to the motor shaft through a fixture with a width of $c$, and the motor is fixed to the frame. Thus, SMA rod 1 cannot move and can only rotate with the motor shaft. The two connection points of SMA rod 1 with the string are subjected to concentrated loads $F_\vartheta$ along the axis of the string. Therefore, SMA rod 1 is divided into three parts: one part with a length of $c$ remains undeformed, while the other two parts undergo deformation symmetrically about the motor axis. Each of these two deformed parts behaves like a cantilever beam subjected to the concentrated load $F_\vartheta$ at its free end. Let $x_\vartheta$ represent the horizontal distance from $A_\vartheta$ to the right end face of the fixture after equilibrium, then $D_\vartheta$ can be expressed as:

$$D_\vartheta = 2x_\vartheta + c \quad (10)$$

Furthermore, by decomposing the tension $F_\vartheta$ at point $A_\vartheta$ into horizontal force $F_{x\vartheta}$ and vertical force $F_{y\vartheta}$, we have:

$$\begin{cases} F_{x\vartheta} = F_z \tan\alpha_\vartheta/2 \\ F_{y\vartheta} = F_z/2 \end{cases} \quad (11)$$

Since the deformation and forces on the left and right deformed parts of SMA rod 1 are symmetrical, we only study the right deformed part. Next, we model the right deformed part of SMA rod 1 as a cantilever beam subjected to the concentrated load $F_\vartheta$ at the free end. Here, the SMA cantilever beam is considered as a straight rod with circular cross-section, where the initial length of the cantilever beam is $l = (D_0 - c)/2$ and the cross-sectional radius is $R$. Due to the much larger length of the SMA cantilever beam compared to its diameter, it can be treated as a slender rod. Thus, the SMA cantilever beam is assumed to satisfy the Euler–Bernoulli theory and Kirchhoff's hypothesis, implying no transverse shear deformation, invariant cross-sections before and after deformation, and cross-sections perpendicular to the axis throughout.

As the SMA cantilever beam exhibits superelastic behavior, it introduces material nonlinearity (variable modulus of elasticity) and geometric nonlinearity (large deformations). Therefore, we establish an accurate mathematical model for the SMA cantilever beam based on the nonlinear large deformation theory of extensible beams (or rods) and considering material nonlinearity. The basic equations of the model include the stress-strain constitutive equation of the SMA material, the geometric equation, and the equilibrium equation. However, the establishment process of this model mainly refers to the method proposed by Shang and Wang. Therefore, this paper only presents the derivation process of the stress-strain constitutive equation of the SMA material due to the use of different SMA materials. As for the geometric equation and the equilibrium equation, only the results are provided, and the detailed process can be referred to in [21].

### 3.2.2 Stress-strain constitutive equation of the SMA material

Due to the complex nonlinear stress-strain behavior of SMA, this paper directly adopts the constitutive relationship of the material fitted from the stress-strain experimental curve of SMA. The SMA material used in the experiment is a superelastic TiNi SMA rod provided by Shenzhen Lianhe Lilong Metal Products Co., Ltd. The chemical composition of the SMA rod is Ti-49.9 wt% Ni, with a phase transition temperature range of 35-65°C, a diameter of 0.9 mm, and an austenite phase composition at room temperature. The stress-strain experimental curve of the SMA material was obtained through uniaxial tensile tests at room temperature. The entire experimental process consists of three stages: the austenite linear elastic deformation stage, the nonlinear stage of stress-induced martensitic phase transformation, and the martensite linear elastic deformation stage. By fitting the experimental curve using the least squares method, the stress-strain experimental curve of the SMA material and the fitting curve are shown in Figure 4. The constitutive equation of stress-strain, fitted with a fifth-order polynomial, is expressed as follows:

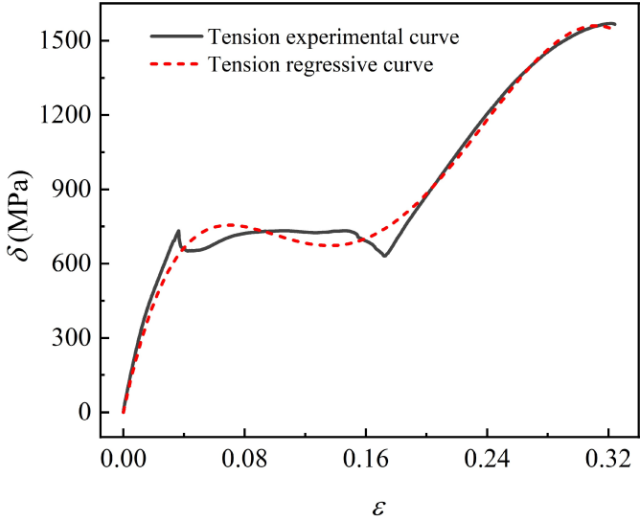

Fig. 4 Stress-strain curve of SMA

$$\sigma = a_1\varepsilon + a_2\varepsilon^2 + a_3\varepsilon^3 + a_4\varepsilon^4 + a_5\varepsilon^5 \quad (12)$$

where: $a_1 = 2\times 10^4$, $a_2 = 2\times 10^4$, $a_3 = 2\times 10^4$, $a_4 = 2\times 10^4$, $a_5 = 2\times 10^4$.

### 3.2.3 Geometric equation

The geometric equations obtained are as follows:

$$dx + \frac{du}{dx}dx = ds\cos\phi$$
$$dw = ds\sin\phi \quad (13)$$

Let the deformation ratio of the axis be $\delta$, then

$$\frac{ds}{dx} = \delta \quad (14)$$

So, the strain of any point on the cross-section can be expressed as

$$\varepsilon = \delta - 1 - y\frac{d\phi}{dx} \quad (15)$$

Here, $x$ is the distance from an arbitrary point C on the rod's axis to the right end face of the fixture. $y$ is the distance from another point P, selected on the cross-section passing through point C, to the rod's axis. $s$ is the arc length of the rod's axis after deformation. $u$ and $w$ denote the axial and lateral displacements of point C, respectively and $\phi$ represents the rotation angle of the corresponding cross-section after deformation. $du$, $dw$, and $d\phi$ denote their respective increments. $dx$ is the axial length of the small rod element before deformation, and $ds$ is the axial length of the small rod element after deformation.

### 3.2.4 Equilibrium equation

The axial force $N(x)$ equivalent to all normal stresses on the cross-section of the SMA cantilever beam is given by:

$$N = \iint \sigma \, dA = \iint \left[a_1\varepsilon + a_2\varepsilon^2 + a_3\varepsilon^3 + a_4\varepsilon^4 + a_5\varepsilon^5\right] dA$$

$$= A_1(\delta-1) + A_2(\delta-1)^2 + A_3(\delta-1)^3 + A_4(\delta-1)^4 + A_5(\delta-1)^5$$

$$+ \left[B_0 + B_1(\delta-1) + B_2(\delta-1)^2 + B_3(\delta-1)^3\right]\left(\frac{d\phi}{dx}\right)^2$$

$$+ \left[C_0 + C_1(\delta-1)\right]\left(\frac{d\phi}{dx}\right)^4 \quad (16)$$

The concentrated moment $M(x)$ due to all normal stresses on the cross-section is given by:

$$M = \iint \sigma y \, dA = \iint \left[a_1\varepsilon + a_2\varepsilon^2 + a_3\varepsilon^3 + a_4\varepsilon^4 + a_5\varepsilon^5\right] y \, dA$$

$$= \left[I_0 + I_1(\delta-1) + I_2(\delta-1)^2 + I_3(\delta-1)^3 + I_4(\delta-1)^4\right]\frac{d\phi}{dx}$$

$$+ \left[J_0 + J_1(\delta-1) + J_2(\delta-1)^2\right]\left(\frac{d\phi}{dx}\right)^3 + K_0\left(\frac{d\phi}{dx}\right)^5 \quad (17)$$

where: $A_i = a_i\pi R^2$ for $i = 1,\ldots,5$; $B_0 = \frac{1}{3}a_2\pi R^4$, $B_1 = \frac{3}{2}a_3\pi R^4$, $B_2 = \frac{3}{2}a_4\pi R^4$, $B_3 = \frac{5}{2}a_5\pi R^4$, $C_0 = \frac{1}{8}a_4\pi R^6$, $C_1 = \frac{5}{8}a_5\pi R^6$, $I_1 = -\frac{1}{4}a_2\pi R^4$, $I_2 = -\frac{3}{4}a_3\pi R^4$, $I_3 = -a_4\pi R^4$, $I_4 = -\frac{5}{4}a_5\pi R^4$, $J_0 = -\frac{1}{8}a_3\pi R^6$, $J_1 = -\frac{1}{2}a_4\pi R^6$, $J_2 = -\frac{5}{4}a_5\pi R^6$, and $K_0 = -\frac{5}{64}a_5\pi R^8$.

The equilibrium equations can be described as follows:

$$\frac{dF_x}{dx} = 0$$
$$\frac{dF_y}{dx} = 0 \quad (18)$$
$$\frac{dM}{dx} = \delta(-F_x\sin\phi + F_y\cos\phi)$$

Where $F_x$, $F_y$, and $M$ represent the internal force components parallel to the coordinate axes and the bending moment on the cross-section, respectively and bending moment on the cross-section, while $dF_x$, $dF_y$, and $dM$ are their corresponding increments. According to the theory of equivalent forces, the axial force $N$ (assuming tension is positive) is expressed as:

$$N = -F_x\cos\phi + F_y\sin\phi \quad (19)$$

### 3.2.5 Basic governing differential equations and boundary conditions

Taking $s$, $u$, $w$, $\phi$, $F_x$, $F_y$, and $M$ as the fundamental unknown functions in the deformation problem for the SMA cantilever beam subjected to concentrated forces in the horizontal and vertical directions at the free end, the basic governing differential equations can be summarized as follows:

$$\frac{du}{dx} - \frac{ds}{dx}\cos\phi + 1 = 0;$$

$$\frac{dw}{dx} - \frac{ds}{dx}\sin\phi = 0;$$

$$A_1\left(\frac{ds}{dx}-1\right) + A_2\left(\frac{ds}{dx}-1\right)^2 + A_3\left(\frac{ds}{dx}-1\right)^3$$
$$+ A_4\left(\frac{ds}{dx}-1\right)^4 + A_5\left(\frac{ds}{dx}-1\right)^5 + \left[B_0 + B_1\left(\frac{ds}{dx}-1\right)\right.$$
$$+ B_2\left(\frac{ds}{dx}-1\right)^2 + B_3\left(\frac{ds}{dx}-1\right)^3\left]\left(\frac{d\phi}{dx}\right)^2 + [C_0\right.$$
$$+ C_1\left(\frac{ds}{dx}-1\right)\left]\left(\frac{d\phi}{dx}\right)^4 + F_x\cos\phi + F_y\sin\phi = 0;\right.$$

$$I_0 + I_1\left(\frac{ds}{dx}-1\right) + I_2\left(\frac{ds}{dx}-1\right)^2 + I_3\left(\frac{ds}{dx}-1\right)^3 \quad (20)$$
$$+ I_4\left(\frac{ds}{dx}-1\right)^4\frac{d\phi}{dx} + \left[J_0 + J_1\left(\frac{ds}{dx}-1\right)\right.$$
$$+ J_2\left(\frac{ds}{dx}-1\right)^2\left]\left(\frac{d\phi}{dx}\right)^3 + K_0\left(\frac{d\phi}{dx}\right)^5 + M = 0;\right.$$

$$\frac{dF_x}{dx} = 0;$$

$$\frac{dF_y}{dx} = 0;$$
$$\frac{dM}{dx} + \frac{ds}{dx}\left(F_x\sin\phi - F_y\cos\phi\right) = 0;$$

When the beam deforms, it does so under certain boundary constraints. Since the the left end of SMA cantilever beam is fixed and the free end is subjected to concentrated forces, the boundary conditions consisting of displacement and force boundaries are as follows:

$$s(0) = 0, u(0) = 0, w(0) = 0, \phi(0) = 0, \quad (21)$$
$$F_x(l) = -F_{x\vartheta}, F_y(l) = F_{y\vartheta}, M(l) = 0$$

Here, $s(0)$, $u(0)$, $w(0)$, and $\phi(0)$ represent the values of $s$, $u$, $w$, and $\phi$ at the fixed end, respectively, while $F_x(l)$, $F_y(l)$, and $M(l)$ represent the values of $F_x$, $F_y$, and $M$ at the free end.

From equations (18), we know that $dF_x = 0$ and $dF_y = 0$, therefore, we can determine that the values of $F_x$ and $F_y$ at the fixed end, denoted as $F_x(0)$ and $F_y(0)$ respectively, are:

$$\begin{cases} F_x(0) + F_x(l) = 0, F_x(0) = F_{x\vartheta} \\ F_y(0) + F_y(l) = 0, F_y(0) = -F_{y\vartheta} \end{cases} \quad (22)$$

According to equation (11), we have

$$\begin{cases} F_x(0) = F_z/2\tan\alpha_\vartheta \\ F_y(0) = -F_z/2 \end{cases} \quad (23)$$

Finally, the boundary conditions for the basic set of governing differential equations become:

$$s(0) = 0, u(0) = 0, w(0) = 0, \phi(0) = 0, \quad (24)$$
$$F_x(0) = F_z/2\tan\alpha_\vartheta, F_y(0) = -F_z/2, M(l) = 0$$

### 3.3 Integrating TSA fundamental model with SMA rod model

The correlation between the SMA rod model and the TSA fundamental model lies in the helix angle $\alpha_\vartheta$, according to equation (5), we have:

$$\tan\alpha_\vartheta = \frac{D_\vartheta + \vartheta r_\vartheta}{X_\vartheta} \quad (25)$$

According to the definition in the SMA cantilever beam model, equation (10) transforms into:

$$D_\vartheta = 2(l + u(l)) + c \quad (26)$$

where $u(l)$ represents the value of $u$ at the free end.

Substituting equation (26) into equation (25), we obtain:

$$\tan\alpha_\vartheta = \frac{2(l + u(l)) + c + \vartheta r}{L_0\cos\alpha_\vartheta + \frac{F_z}{2K}} \quad (27)$$

By combining equations (20), (27), and the boundary condition equation (24), all variables of the entire mechanism can be solved. Since $y_\vartheta = w(l)$ and $y_1 = w(l)_{\vartheta=\pi}$, equation (3) be expressed as:

$$\Delta X_\vartheta = X_1 - X_\vartheta - 2(y_\vartheta - y_1)$$
$$= L_0(\cos\alpha_{\vartheta=\pi} - \cos\alpha_\vartheta) - 2(w(l)_\vartheta - w(l)_{\vartheta=\pi}) \quad (28)$$

where $w(l)_\vartheta$ and $w(l)_{\vartheta=\pi}$ represent the respective values of $w$ at the free end when the motor rotates by an angle of $\vartheta$ and $\pi$.

Finally, the TR of the proposed TSA-based CVT mechanism is given by:

$$TR = \frac{1}{\partial(\Delta X_\vartheta)/\partial\vartheta} \quad (29)$$

## 4 Numerical simulation and discussion

### 4.1 Numerical simulation method

The solution of the proposed TSA-based CVT mechanism involves the joint resolution of the basic control differential equation group for SMA cantilever beam, along with the fundamental equation group for TSA. Solving the basic control differential equation group for the SMA cantilever beam entails resolving a highly nonlinear implicit differential equation group with seven unknown functions, posing a boundary value problem. Moreover, the boundary conditions dynamically shift in response to the twisted state of the strings, linked by the helix angle $\alpha_\vartheta$. Currently, there is no analytical solution for the boundary value problem of the nonlinear implicit differential equation group, thus rendering no analytical solution for the entire TSA-based CVT mechanism. Additionally, with different motor rotation angles $\vartheta$ and varying numbers of strings twists, the helix angle $\alpha_\vartheta$ differs, leading to different boundary conditions for solving the nonlinear implicit differential equation group of the SMA cantilever beam. Therefore, we adopt a solution procedure as shown in Figure 5 and utilize numerical methods combining the shooting method, Newton–Raphson iteration method, and Runge-Kutta method to solve the boundary value problem of the nonlinear implicit differential equation group. We first assign a motor rotation angle $\vartheta$, and the mechanism will have a corresponding helix angle $\alpha_\vartheta$ in the equilibrium state. As this helix angle $\alpha_\vartheta$ must satisfy both the basic control differential equation group for SMA rod deformation and the fundamental equation group for TSA, we adjust the given value continuously until it simultaneously satisfies both equation groups, thereby obtaining the solution for all unknown quantities of the entire mechanism under the given motor rotation angle. By providing a continuously changing motor rotation angle $\vartheta$, we can ultimately calculate the corresponding $\Delta X_\vartheta$ and TR of the TSA-based CVT mechanism as the motor rotation angle $\vartheta$ changes.

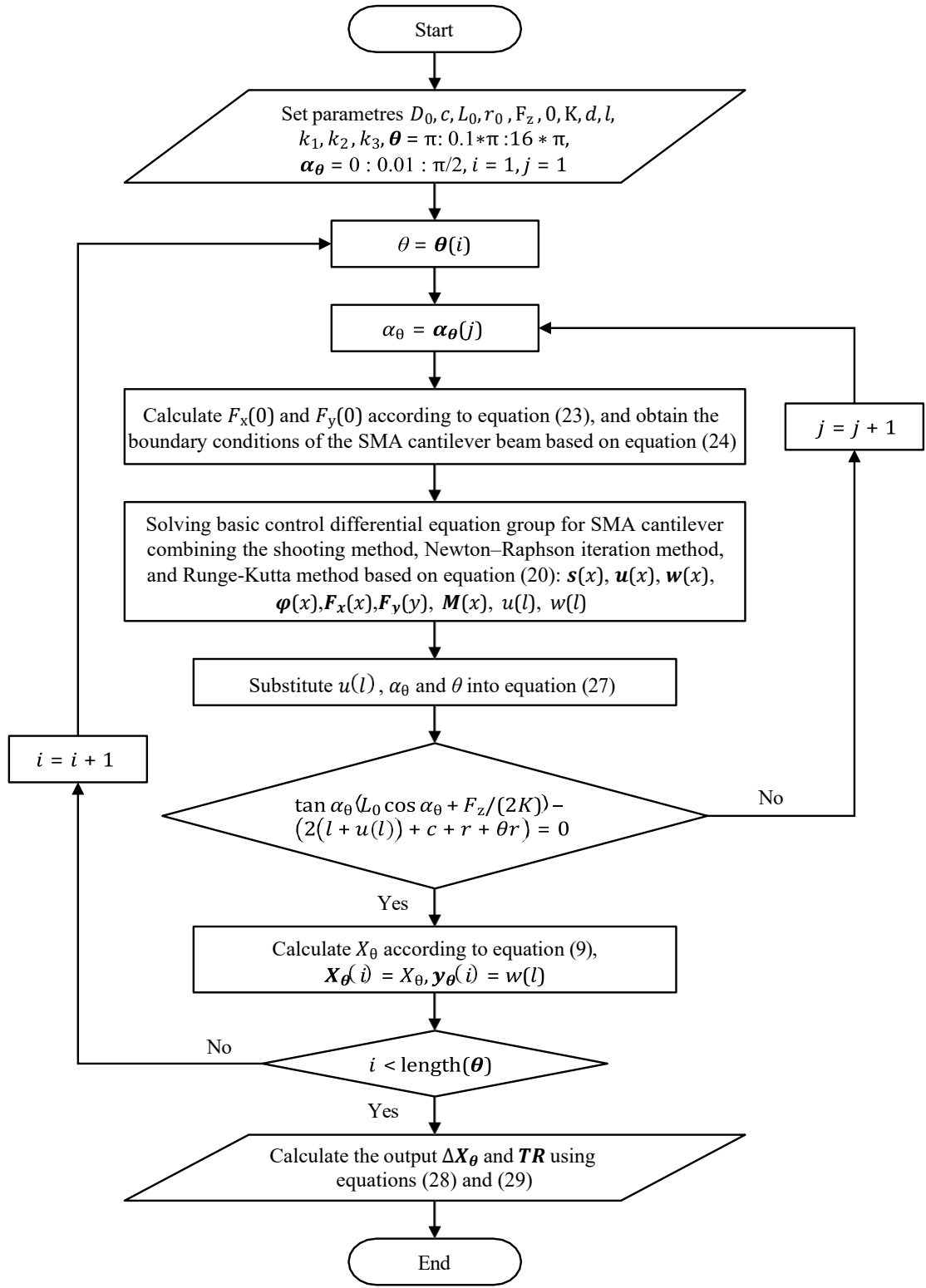

Fig. 5 Flowchart of numerical solution algorithm.

## 4.2 Results and discussion

Table 1 The parameters employed in the theoretical modeling of the TSA-based CVT mechanism.

| $L_0$ (m) | $D_0$ (m) | $c$ (m) | $r_0$ (m) | $R$ (m) | $K$ (kN/m) | $k_1$ (1/kN) | $k_2$ | $k_3$ |
|---|---|---|---|---|---|---|---|---|
| 0.097 | 0.032 | 0.006 | 0.0015 | 0.0005 | 1.37 | 6522 | 0.127 | 0.133 |

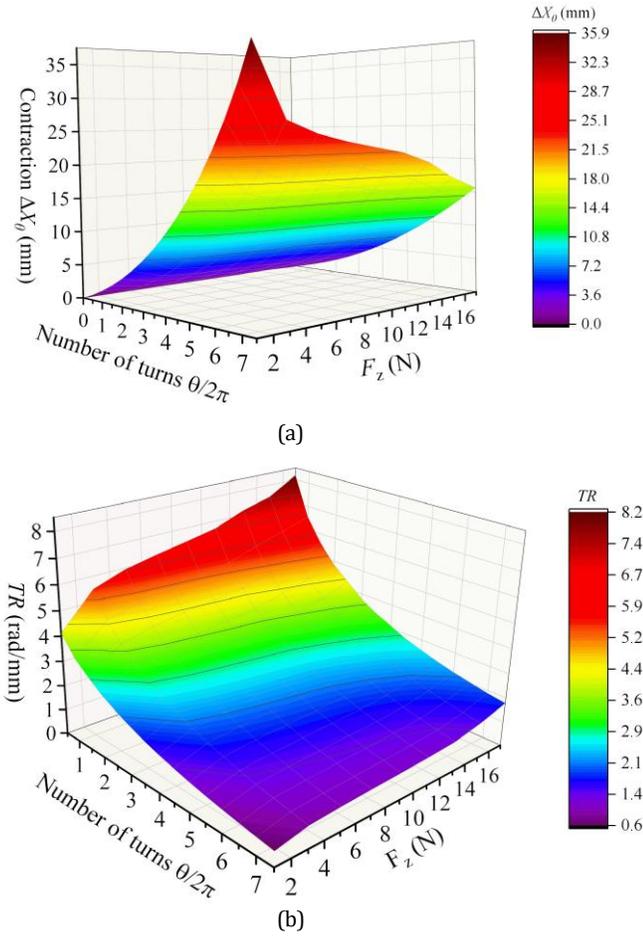

(a)

(b)

Fig. 6 Results of the theoretical model calculations for the TSA-based CVT mechanism: Variations in the contraction $\Delta X_\vartheta$ (a) and $TR$ (b) with changes in load and motor's number of turns.

Using the numerical approach described above, the complete system of equations is solved. The variations in the contraction $\Delta X_\vartheta$ and the $TR$ of the TSA-based CVT mechanism under different loads $F_z$ with respect to the number of motor turns are calculated. The results are shown in Figure 6. Additionally, Table 1 lists the parameters used in the theoretical model.

It can be observed that as the motor's number of turns increases, $\Delta X_\vartheta$ gradually increases while $TR$ gradually decreases, which is consistent with the findings of the Ryu team[19,24]. Additionally, for the same motor's number of turns, as the load increases, $\Delta X_\vartheta$ gradually decreases while $TR$ gradually increases. Under a load of 1 N, after the motor continuously rotates 7.5 turns from the initial condition of half a pre-twist, the contraction $\Delta X_\vartheta$ of the TSA-based CVT mechanism is 35.89 mm, whereas under a load of 17 N, the contraction $\Delta X_\vartheta$ is 15.64 mm, reducing to 43.58% of the original value. With an initial pre-twist of half a turn, when the load increases from 1 N to 17 N, $TR$ increases from 4.13 rad/mm to 8.19 rad/mm, which is 1.98 times the original value. With the motor continuously rotating 7.5 turns from the initial condition of half a pre-twist, $TR$ increases from 0.60 rad/mm to 1.78 rad/mm under the corresponding load conditions, which is 2.97 times the original value. These results indicate that for lighter load, the TSA-based CVT mechanism exhibits a lower $TR$, allowing for faster contraction speed, whereas for heavier load, the TSA-based CVT mechanism shows a higher $TR$, resulting in slower contraction speed but enhanced load capacity. Moreover, under the same load variation condition, the $TR$ of the TSA-based CVT mechanism demonstrates more pronounced TR sensitivity to load variations as the number of turns increases. The contraction $\Delta X_\vartheta$ and $TR$ both exhibit a continuous variation trend with the motor's number of turns and the load changes. Therefore, the effectiveness of the adaptive load-variable TR of the designed TSA-based CVT mechanism is theoretically validated.

To explore the impact of parameter variations on the TR in the TSA-based CVT mechanism, the main parameters considered in the mechanism model are the initial length of the SMA rod $D_0$, the radius of the SMA rod $R$, the initial length of the twisted string $L_0$, the initial diameter of the twisted string $r_0$, and the stiffness coefficient of the twisted string $K$. The effects of these parameters on the contraction $\Delta X_\vartheta$ and $TR$ of the TSA-based CVT mechanism will be explored individually. All calculations are performed under a load force $F_z$ of 7.5 N. Each calculation varies only the parameter under investigation, while the other parameters remain as specified in Table 1.

The effects of different initial lengths $D_0$ of the SMA rod on the contraction $\Delta X_\vartheta$ and $TR$ of the TSA-based CVT mechanism are shown in Figure 7. It can be observed that for the same motor's number of turns, a larger initial length $D_0$ results in a greater contraction $\Delta X_\vartheta$ and a smaller $TR$. Under conditions where the initial lengths of the SMA rod $D_0$ are 8 mm, 14 mm, 20 mm, 26 mm, and 32 mm, the contraction $\Delta X_\vartheta$ after the motor continuously rotates 7.5 turns from an initial pre-twist of half a turn are 11.77 mm, 14.60 mm, 17.63 mm, 20.85 mm, and 23.83 mm, respectively. As the initial length $D_0$ of the SMA rod increases by equal intervals, the contraction $\Delta X_\vartheta$ increases by 2.83 mm, 3.04 mm, 3.21 mm, and 2.98 mm. This indicates that increasing the initial length $D_0$ of the SMA rod leads to a larger contraction $\Delta X_\vartheta$ for the TSA-based CVT mechanism. However, as $D_0$ increases, the rate of increase in $\Delta X_\vartheta$ first increases and then decreases. This is because, for traditional fixed-offset TSAs, an increase in $D_0$ generally results in an increase in $\Delta X_\vartheta$. However, due to bending deformation of the SMA rod under load, a longer SMA rod deforms more easily under the same load, making the horizontal distance between twisted strings shorter in the loaded state, which reduces $\Delta X_\vartheta$. The combined effect of these two factors presents the final trend in $\Delta X_\vartheta$. Therefore, simply increasing the initial length $D_0$ of the SMA rod does not always lead to a greater $\Delta X_\vartheta$ and reduced $TR$.

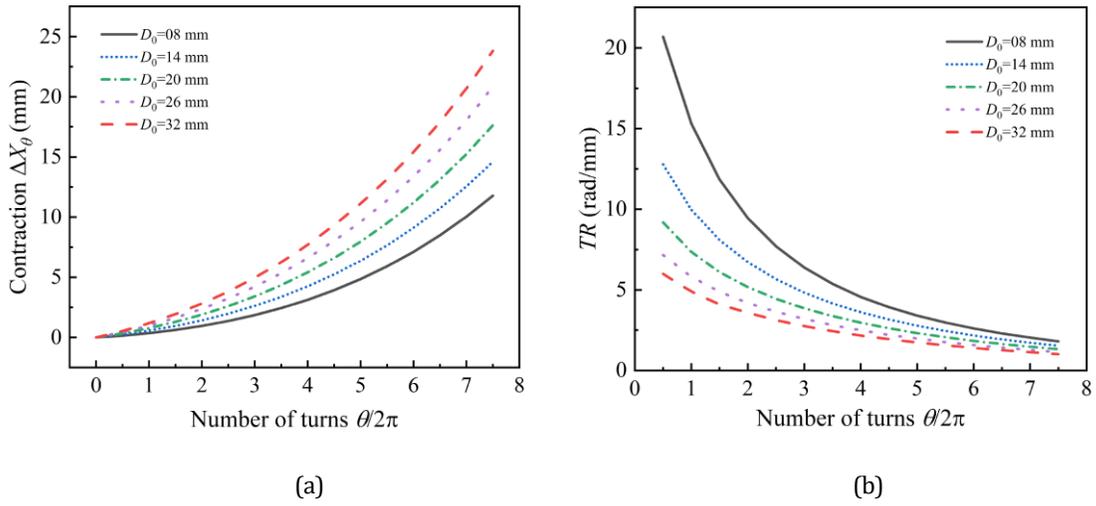

Fig. 7 Variation in the contraction $\Delta X_\vartheta$ (a) and $TR$ (b) of the TSA-based CVT mechanism under different initial lengths $D_0$ of the SMA rod.

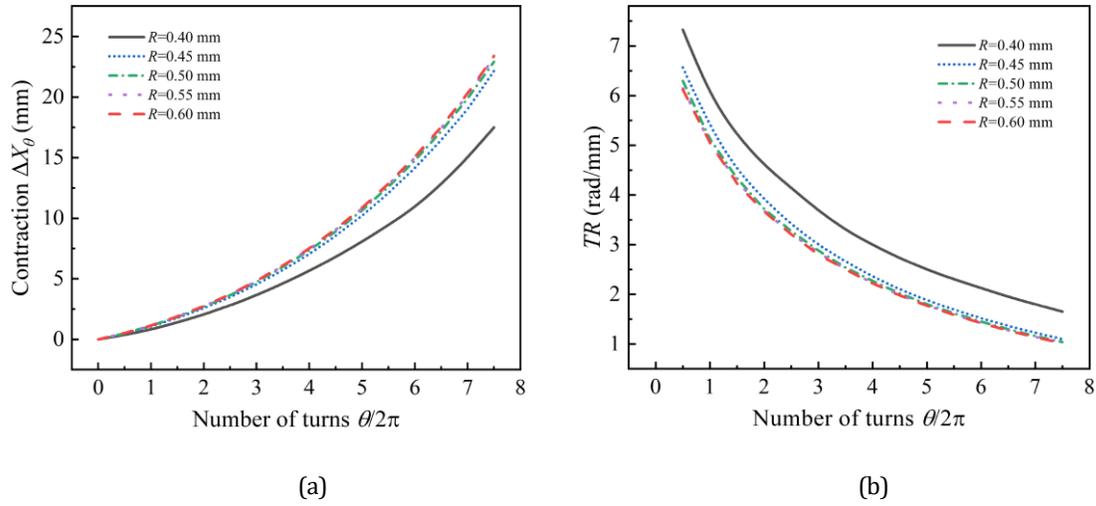

Fig. 8 Variation in the contraction $\Delta X_\vartheta$ (a) and $TR$ (b) of the TSA-based CVT mechanism under different radius $R$ of the SMA rod.

The effects of different radius $R$ of the SMA rod on the contraction $\Delta X_\vartheta$ and $TR$ of the TSA-based CVT mechanism are shown in Figure 8. It can be observed that, for the same motor's number of turns, as the radius $R$ of the SMA rod increases, the contraction $\Delta X_\vartheta$ gradually increases while the $TR$ gradually decreases. However, after a certain radius $R$ is reached, further increases in $R$ result in negligible changes in $\Delta X_\vartheta$ and $TR$. For example, when $R$ increases from 0.4 mm to 0.45 mm, $\Delta X_\vartheta$ increases from 17.50 mm to 22.16 mm, showing a change of 4.66 mm. However, when $R$ increases from 0.45 mm to 0.50 mm, $\Delta X_\vartheta$ only increases from 22.16 mm to 22.92 mm, showing a smaller change of 0.76 mm. For $R$ values of 0.55 mm and 0.60 mm, $\Delta X_\vartheta$ remains almost constant. This is because, under a load of 7.5 N, a smaller radius $R$ of SMA rod has lower bending strength and undergoes significant bending deformation, reducing the horizontal distance between twisted strings, resulting in smaller $\Delta X_\vartheta$ and larger $TR$.

As the radius $R$ of the SMA rod increases, its bending strength increases, causing less bending deformation under the same load, and thus a smaller reduction in the horizontal distance between twisted strings, leading to larger $\Delta X_\vartheta$ and smaller $TR$. When the radius $R$ reaches a certain point, the SMA rod undergoes minimal deformation, and under a 7.5 N load, the TSA-based CVT mechanism behaves similarly to a traditional fixed-offset TSA. Further increases in $R$ will no longer affect $\Delta X_\vartheta$ and $TR$. Therefore, it is important to select an appropriate SMA rod radius $R$ based on the desired load, to ensure effective adjustment of $\Delta X_\vartheta$ and $TR$ in the TSA-based CVT mechanism.

The effects of different initial lengths $L_0$ of the twisted string on the contraction $\Delta X_\vartheta$ and $TR$ of the TSA-based CVT mechanism are shown in Figure 9. It can be observed that, for the same motor's number of turns, a larger initial length $L_0$ results in a smaller contraction $\Delta X_\vartheta$ and a larger $TR$. Under initial lengths $L_0$ of 70 mm, 110 mm, 150 mm, 190 mm, and 230 mm for the the twisted string, the contraction $\Delta X_\vartheta$ after the motor continuously rotates 7.5 turns from an initial pre-twist of half a turn are 38.67 mm, 19.38 mm, 13.39 mm, 10.33 mm, and 8.44 mm, respectively. As the initial length $L_0$ increases by equal intervals, the contraction $\Delta X_\vartheta$ decreases by 19.29 mm, 5.99 mm, 3.06 mm, and 1.89 mm, respectively. This indicates that increasing the initial length $L_0$ of the twisted string reduces the contraction $\Delta X_\vartheta$ of the TSA-based CVT mechanism, but the rate of reduction decreases as $L_0$ increases. When the initial length $L_0$ is between 70 mm and 110 mm with $D_0$ set to 30 mm and a load force of 7.5 N, the mechanism's $\Delta X_\vartheta$ and $TR$ are notably sensitive to changes in $L_0$. This provides important reference for applications requiring specific $\Delta X_\vartheta$ while aiming to minimize mechanism size.

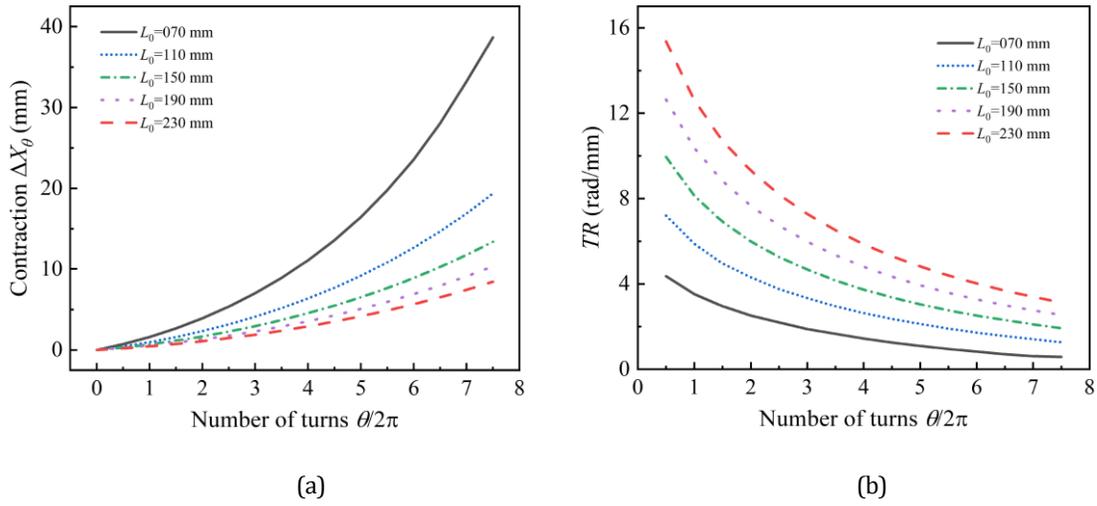

Fig. 9 Variation in the contraction $\Delta X_\vartheta$ (a) and $TR$ (b) of the TSA-based CVT mechanism under different initial length $L_0$ of the twisted string.

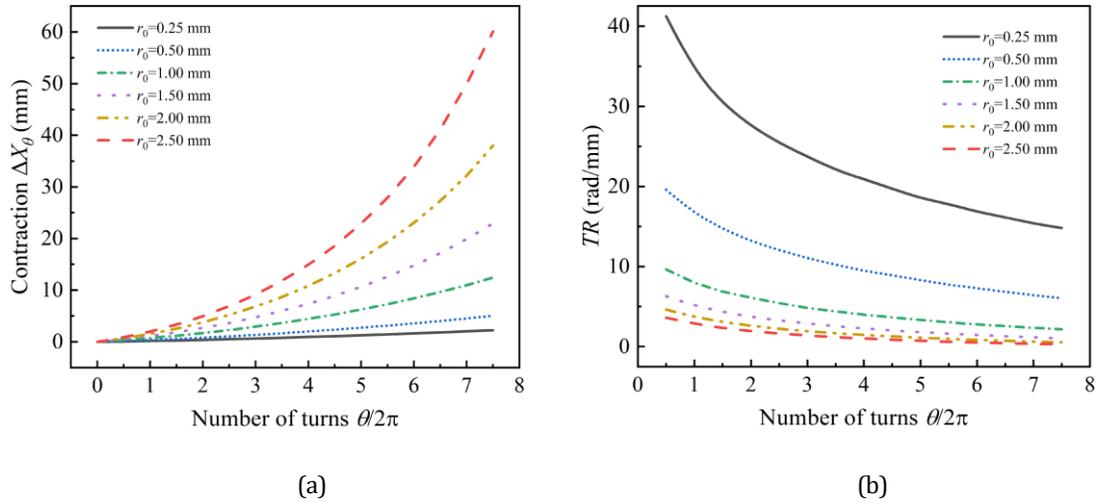

Fig. 10 Variation in the contraction $\Delta X_\vartheta$ (a) and $TR$ (b) of the TSA-based CVT mechanism under different initial diameter $r_0$ of the twisted string.

The effects of different initial diameter $r_0$ of the twisted string on the contraction $\Delta X_\vartheta$ and $TR$ of the TSA-based CVT mechanism are shown in Figure 10. It can be observed that, for the same motor's number of turns, a larger twisted string diameter $r_0$ results in a greater contraction $\Delta X_\vartheta$ and a smaller $TR$. For twisted string diameter $r_0$ of 0.25 mm, 0.5 mm, 1.0 mm, 1.5 mm, 2.0 mm, and 2.5 mm, the contraction $\Delta X_\vartheta$ after the motor continuously rotates 7.5 turns from an initial pre-twist of half a turn are 2.24 mm, 5.02 mm, 12.42 mm, 22.92 mm, 38.00 mm, and 60.11 mm, respectively. As the twisted string diameter $r_0$ increases by equal intervals, the contraction $\Delta X_\vartheta$ changes by 2.78 mm, 7.40 mm, 10.50 mm, 15.08 mm, and 22.11 mm, respectively. This indicates that increasing the twisted string diameter $r_0$ results in an increase in the contraction $\Delta X_\vartheta$ of the TSA-based CVT mechanism, with the increase being more pronounced as $r_0$ gets larger. Changing the twisted string diameter $r_0$ has a significant impact on the contraction $\Delta X_\vartheta$ and $TR$. For example, increasing $r_0$ from 0.25 mm to 2.5 mm results in a 26.83-fold increase in $\Delta X_\vartheta$. Therefore, $\Delta X_\vartheta$ and $TR$ are highly sensitive to changes in the twisted string diameter $r_0$. Since increasing $r_0$ has a relatively small effect on the overall mechanism size, adjusting $r_0$ is a preferred method for achieving the required $\Delta X_\vartheta$ while maintaining a compact structure.

The effects of different twisted string stiffness coefficients $K$ on the contraction $\Delta X_\vartheta$ and $TR$ of the TSA-based CVT mechanism are shown in Figure 11. It can be observed that, for the same motor's number of turns, an increase in the twisted string stiffness coefficient $K$ results in a gradual increase in the contraction $\Delta X_\vartheta$

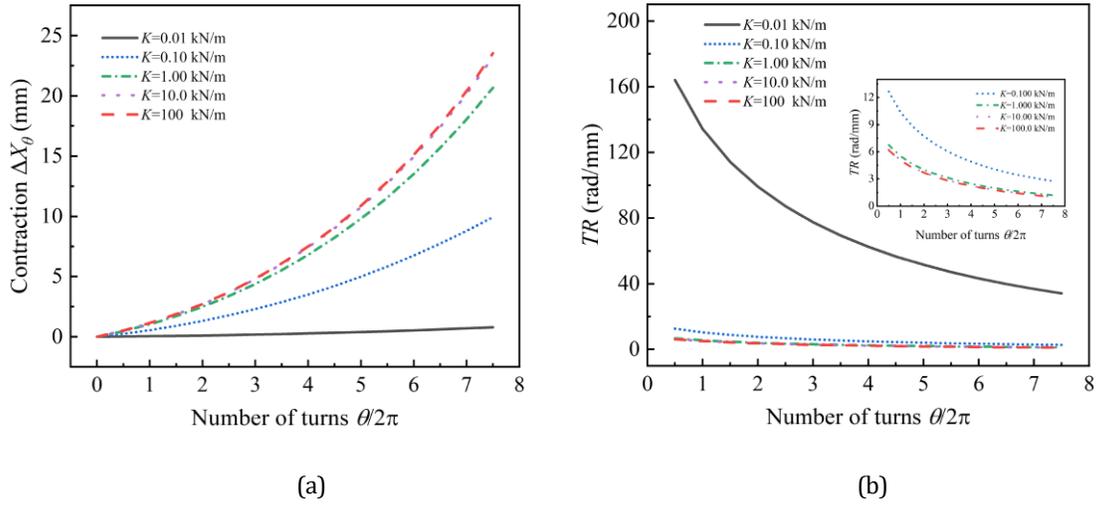

Fig. 11 Variation in the contraction $\Delta X_\vartheta$ (a) and $TR$ (b) of the TSA-based CVT mechanism under different twisted string stiffness coefficients $K$

and a gradual decrease in $TR$. However, after $K$ reaches a certain level, further increases in $K$ result in minimal changes in $\Delta X_\vartheta$ and $TR$. For instance, when $K$ increases from 0.01 kN/m to 0.1 kN/m, $\Delta X_\vartheta$ increases from 0.79 mm to 9.94 mm, a change of 9.15 mm. However, when $K$ increases from 1.0 kN/m to 10.0 kN/m, $\Delta X_\vartheta$ increases from 20.68 mm to 23.24 mm, a much smaller change of 2.55 mm. At $K$ of 100 kN/m, $\Delta X_\vartheta$ is nearly the same as at $K$ of 10 kN/m. This behavior is due to the fact that, under a load of 7.5 N, a twisted string with a smaller $K$ has lower tensile strength and undergoes significant elongation, increasing its length. As the motor rotates more, the string contracts more, but the elongation caused by the tensile force also increases, counteracting part of the contraction due to the increase in rotation, resulting in smaller $\Delta X_\vartheta$ and larger $TR$. As $K$ increases, the string's tensile strength improves, causing less elongation under the same load. This reduces the counteracting effect on $\Delta X_\vartheta$, resulting in larger $\Delta X_\vartheta$ and smaller $TR$. When $K$ reaches a certain level, the string undergoes negligible deformation, and further increases in $K$ do not significantly affect $\Delta X_\vartheta$ and $TR$. Given that TSA-based CVT mechanism applications often require a significant contraction, strings with a higher stiffness coefficient $K$ are generally chosen to meet this requirement.

## 5 Experimental analysis

In this section, an experimental setup was designed to test the performance of the TSA-based CVT mechanism and validate the effectiveness of the mechanism through experiments. Additionally, the validity of the previously established theoretical model will be verified through comparative analysis between experimental measurements and theoretical predictions of the contraction $\Delta X_\vartheta$.

### 5.1 Experimental setup

An performance testing platform of TSA-based CVT mechanism was constructed, with its physical and schematic diagrams shown in Figure 12. The testing platform mainly includes the following components: a DC motor with an encoder (Maxon ENX16), a motor driver (RMDS-405), a computer, a pair of superelastic SMA rods (SMA Rod 1 and SMA Rod 2), two twisted strings, linear guide rail and slider (Hiwin, MG), a linear variable differential transformer (LVDT, W-DC, 20 cm stroke), a load, and a frame. The DC motor is horizontally mounted on the frame and driven by the RMDS-405 motor driver. The load and LVDT are securely fixed to the linear guide rails. Parameters consistent with those given in Table 1 were used for the experiment. Two superelastic nickel-titanium alloy rods with a diameter of 1 mm and a length of 32 mm were used for SMA Rod 1 and SMA Rod 2. The material is the same as that used in the uniaxial tensile test stress-strain curves. SMA Rod 1 is connected to the motor output shaft at its midpoint with a 6 mm wide fixed clamp, while SMA Rod 2 is similarly mounted on the slider at its midpoint with a 6 mm wide fixed clamp. SMA Rod 1 and SMA Rod 2 are installed parallel to each other. Two braided ultrahigh molecular weight polyethylene (UHMWPE) fiber strings were used as twisted strings, each with a diameter of 1.5 mm, an initial length of 97 mm (without external load), and a stiffness coefficient of 1.37 kN/m. The twisted strings are fixed to both ends of SMA Rod 1 and SMA Rod 2 with clamps, initially parallel to the motor shaft. To prevent sudden movement of the load, the twisted strings were pre-twisted by half a turn. The motor then drives SMA Rod 1 to continuously rotate 7.5 turns forward and 7.5 turns backward at a constant speed of 12.5 rpm/min under various loads. The load moves in a reciprocating linear motion constrained by the linear guide rails, and the movement of the load is recorded by the LVDT.

### 5.2 Experimental Validation

The performance testing results of the proposed TSA-based CVT mechanism are shown in Figure 13. Figure 13(a) presents the experimental and theoretical contraction $\Delta X_\vartheta$ of the TSA-based CVT mechanism under different loads (0.1 kg, 0.5 kg, and 1.5 kg) as a function of the motor's number of turns $\vartheta/2\pi$ at a constant speed of 12.5 rpm/min. It can be observed that as the motor's number of turns increases, the contraction $\Delta X_\vartheta$ gradually increases. Under the same motor's number of turns, the contraction decreases with increasing load. The TSA-based CVT mechanism produces maximum contraction of 34.9 mm, 22.1 mm, and 16.64 mm for loads of 0.1 kg, 0.5 kg, and 1.5 kg, respectively. The contraction $\Delta X_\vartheta$ for a 0.5 kg load is 63.3% of that for a 0.1 kg load, and for a 1.5 kg load, $\Delta X_\vartheta$ is 75.3% of that for a 0.5 kg load, and 47.7% of that for a 0.1 kg load. The TSA-based CVT mechanism shows greater sensitivity to load changes from 0.1 kg to 0.5 kg compared to changes from 0.5 kg to 1.5 kg.

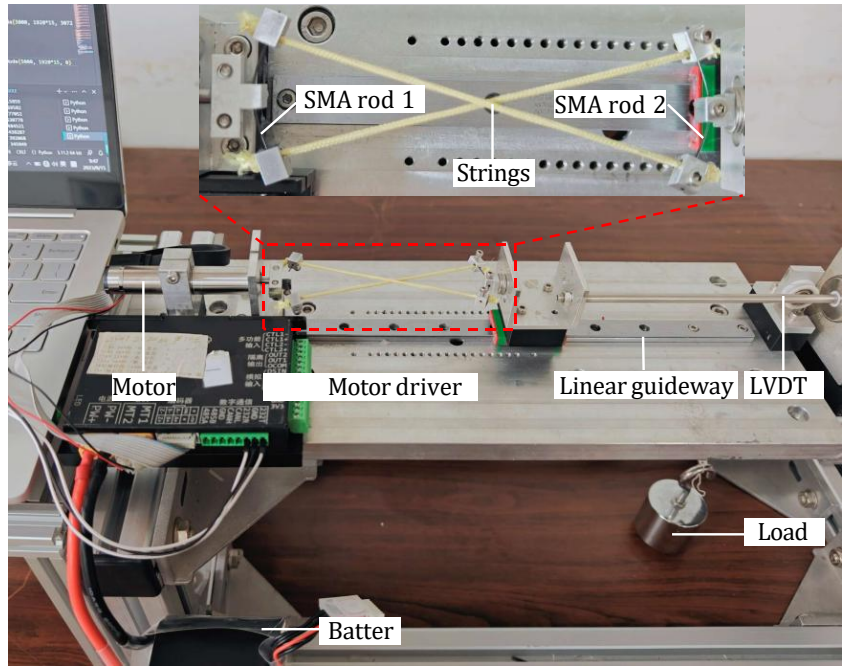

(a)

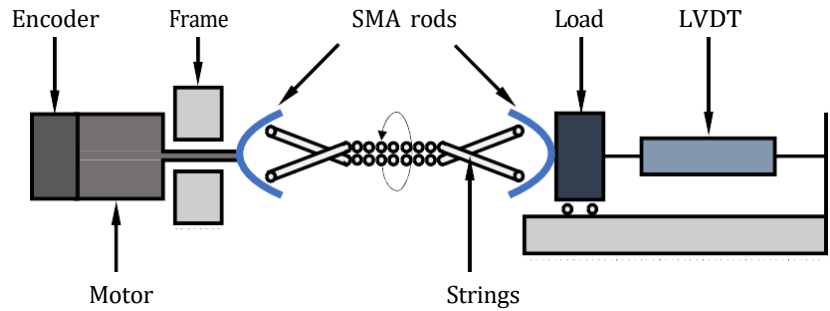

(b)

Fig. 12 The test platform for the properties of TSA-based CVT mechanism: (a) physical diagram; (b) schematic diagram

Figure 13(b) and (c) show the experimental and theoretical average speed and average TR ($AVETR$, the ratio of the motor speed to the average speed of the driven load) of the TSA-based CVT mechanism under different loads at motor's number of turns $\vartheta/2\pi$ of 3.75 and 7.5. It can be seen that the average contraction speed is lower for higher loads and higher for lower loads, whereas the $AVETR$ is higher for higher loads and lower for lower loads. At $\vartheta/2\pi$ of 3.75, the average speed for a 1.5 kg load is 46.8% of that for a 0.1 kg load, and at $\vartheta/2\pi$ of 7.5, the average speed for a 1.5 kg load is 47.7% of that for a 0.1 kg load. For the $AVETR$, at $\vartheta/2\pi$ of 3.75, the $AVETR$ for a 1.5 kg load is 2.13 times that for a 0.1 kg load, and at $\vartheta/2\pi$ of 7.5, the $AVETR$ for a 1.5 kg load is 2.1 times that for a 0.1 kg load. Therefore, for smaller load, the TSA-based CVT mechanism exhibits a smaller $AVETR$, resulting in a faster contraction speed and lower load capacity. Conversely, for larger load, the TSA-based CVT mechanism shows a larger $AVETR$, leading to a slower contraction speed but higher load capacity.

The results above can be explained with reference to Figure 14, which shows snapshots of the TSA-based CVT mechanism under different loads (0.1 kg, 0.5 kg, and 1.5 kg) at both the initial state (Figure 14(a)) and after 7.5 turns from the initial state (Figure 14(b)). In the initial state, when the load is 0.1 kg, the SMA rods experience minimal deformation, with the distance $D_\vartheta$ between the ends of the SMA rods and the twisted strings being 32 mm. As the motor rotates, $D_\vartheta$ remains nearly constant, so the SMA rods operate under a fixed offset of 32 mm, resulting in a faster contraction speed. When the load is increased to 0.5 kg, the SMA rods exhibit slight deformation in the initial state, with $D_\vartheta$ decreasing from 32 mm to 31 mm. As the motor rotation increases, the tension in the strings rises, causing the SMA rods to bend progressively. This bending reduces $D_\vartheta$ and increases the downward displacement $\Delta y_\vartheta$ of the connection points. After 7.5 rotations, $D_\vartheta$ decreases to 28.9 mm. According to equation (3), the contraction of the load is determined by the combined effects of load rise due to string contraction ($X_1 - X_\vartheta$) and load drop due to SMA rod bending ($y_\vartheta - y_1$). A larger $X_1 - X_\vartheta$ results in faster load contraction, while a larger $y_\vartheta - y_1$ results in slower load contraction. As $D_\vartheta$ decreases, the string contraction slows down, making $X_1 - X_\vartheta$ smaller. Therefore, the combined effect of decreasing $D_\vartheta$ and increasing $y_\vartheta - y_1$ reduces the load rise speed. This effect is more pronounced with higher loads. For a 1.5 kg load, the SMA rods are significantly bent in the initial state, reducing $D_\vartheta$ to 29.2 mm, which further decreases to 25.9 mm after 7.5 rotations, resulting in a slower contraction speed. Thus, the experimental results confirm that the proposed TSA-based CVT mechanism can automatically adjust the $TR$ based on external load change, effectively meeting the mechanism's requirements for varying loads.

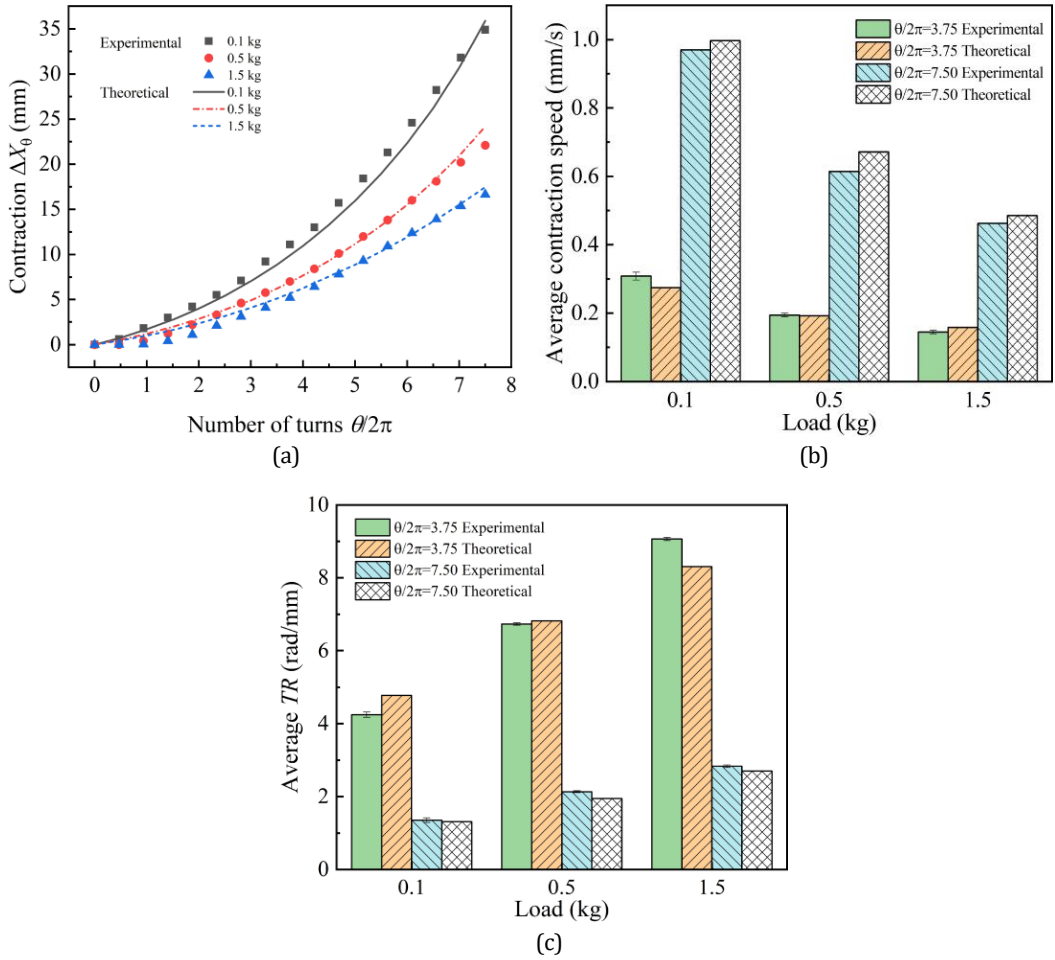

Fig. 13 Comparison of theoretical and experimental results for TSA-based CVT mechanism under different loads: (a) change in contraction $\Delta X_\vartheta$ with motor's number of turns $\vartheta/2\pi$; (b) average contraction speed at motor's number of turns $\vartheta/2\pi$ of 3.75 and 7.5; (c) average TR ($AVETR$) at motor's number of turns $\vartheta/2\pi$ of 3.75 and 7.5.

In addition, Figure 13 demonstrates good agreement between the established theoretical model and experimental results across all loading conditions, except for some initial twisting phase deviation observed under the 1.5 kg load. This discrepancy may be attributed to interstitial gaps within the braided string structure, which are progressively eliminated during the twisting process - particularly under higher load conditions where the gap reduction effect becomes more pronounced. However, there is Overall, the model effectively calculates the load contraction $\Delta X_\vartheta$, contraction speed, and $TR$ of the TSA-based CVT mechanism. This model serves as a valuable theoretical tool for analyzing, optimizing, and controlling the TSA-based CVT mechanism.

## Conclusions

This paper presented the design, comprehensive modeling, and experimental validation of a novel TSA-based CVT mechanism utilizing the superelasticity of rods. The proposed mechanism adjusts the distance between twisted strings via deformation of SMA rods in response to external loads, enabling continuous adaptation of TR to varying load conditions. The proposed mechanism addresses the limitation of fixed TR in conventional TSA systems, offering a lightweight, simple-structured, cost-effective, and highly adaptable solution. In addition, a high-fidelity theo-retical model was developed, simultaneously incorporating three critical nonlinearities: the SMA's material nonlinearity (variable elastic modulus), geometric nonlinearity (large deformations), and axial deformation dynamics. This complex boundary-value problem, governed by highly nonlinear implicit differential equations, was solved numerically using a combined shooting method, Newton-Raphson iteration, and Runge-Kutta method. Theoretical and experimental results validate the mechanism's ability to autonomously adjust TR based on varying loads. The mechanism's theoretical model aligns closely with experimental data, and parameters analysis provided key insights for optimizing performance on contraction $\Delta X_\vartheta$ and $TR$. This mechanism effectively expands the operational versatility of TSAs, making it promising for applications in robotic hands, exoskeletons, and soft robotic systems where adaptive speed-force tradeoffs are critical. Future work will focus on control strategy development and integration into functional robotic system.

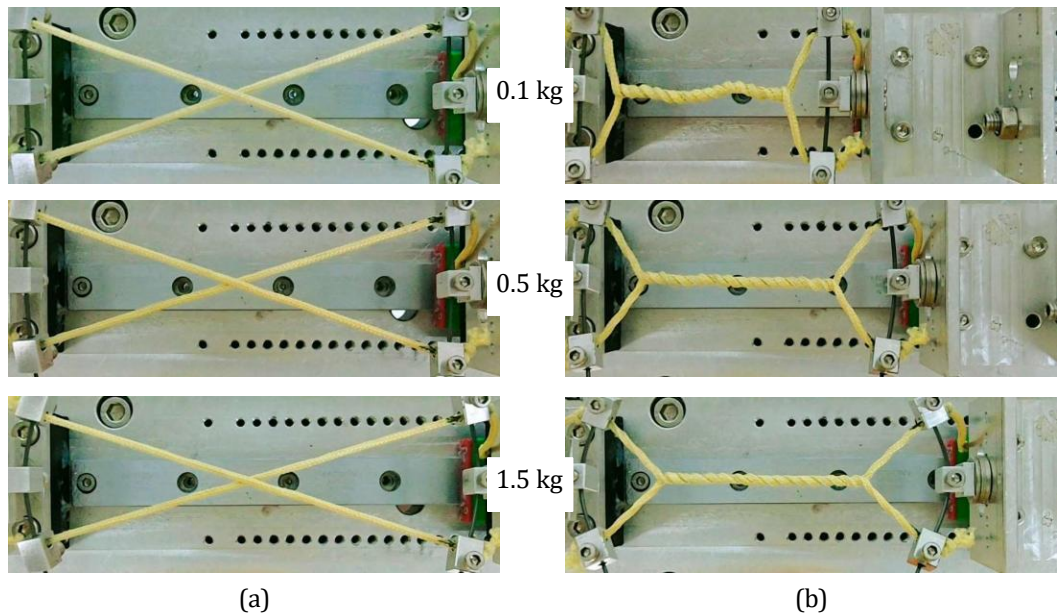

Fig. 14 Snapshot of TSA-based CVT mechanism carrying workloads of 0.1 kg, 0.5 kg, and 1.5 kg. (a) In the initial state for pre-twisted half a turn with load. (b) After being twisted 7.5 turns from the initial state.

## Author contributions

Conceptualization, C.X.; methodology, C.X.; validation, S.D.; formal analysis, C.X.; investigation, C.X.; writing—original draft preparation, C.X. and S.D.; writing—review and editing, C.X., S.D and X.W.; All authors have read and agreed to the published version of the manuscript.

## References

1. T. Würtz, C. May, B. Holz, C. Natale, G. Palli and C. Melchiorri, 2010 IEEE/ASME International Conference on Advanced Intelligent Mechatronics, pp. 1215–1220.
2. T. Sonoda and I. Godler, 2010 IEEE/RSJ International Conference on Intelligent Robots and Systems, 2012, pp. 2733–2738.
3. J. Zhang, J. Sheng, C. T. O'Neill, C. J. Walsh, R. J. Wood, J.-H. Ryu, J. P. Desai and M. C. Yip, *IEEE Transactions on Robotics*, 2019, **35**, 761–781.
4. H. Singh, D. Popov, I. Gaponov and J.-H. Ryu, *Mechanism and Machine Theory*, 2016, **100**, 205–221.
5. C. Xu, S. Dong, Y. Ma, J. Zhan, Y. Wang and X. Wang, *Journal of Bionic Engineering*, 2024, **21**, 1174–1190.
6. M. F. Rahman, K. Zhang and G. Herrmann, Towards Autonomous Robotic Systems: 20th Annual Conference, 2019, pp. 489–492.
7. D. Popov, I. Gaponov and J.-H. Ryu, *IEEE/ASME Transactions on Mechatronics*, 2016, **22**, 865–875.
8. I. W. Park and V. SunSpiral, *2014 14th International Conference on Control, Automation and Systems (Iccas 2014)*, 2014, 1331–1338.
9. A. Baker, C. Foy, S. Swanbeck, R. Konda and J. Zhang, 2023 IEEE/ASME International Conference on Advanced Intelligent Mechatronics (AIM), pp. 643–648.
10. D. Bombara, R. Konda, E. Chow and J. Zhang, 2022 American Control Conference (ACC), 2022, pp. 4365–4370.
11. J. Jang, Y.-U. Song and J.-H. Ryu, *IEEE Robotics and Automation Letters*, 2022, **7**, 2605–2612.
12. S. H. Jeong, Y. J. Shin and K.-S. Kim, *IEEE/ASME Transactions on Mechatronics*, 2017, **22**, 2790–2801.
13. H. Yamada, 2012 IEEE International Conference on Robotics and Automation, 2012, pp. 1982–1987.
14. X. Chen, P. Hang, W. Wang and Y. Li, *Mechanism and Machine Theory*, 2017, **107**, 13–26.
15. Y. J. Shin, K.-H. Rew, K.-S. Kim and S. Kim, 2013 IEEE International Conference on Robotics and Automation, pp. 2759–2764.
16. Z. Lu, R. Wang, Y. Xiao, T. Liu, C. Liu and H. Zhao, *Advanced Robotics*, 2023, **37**, 1520–1531.

17 S. Kim, J. Sim and J. Park, *IEEE Robotics and Automation Letters*, 2020, **5**, 5477–5484.
18 W. Shin, S. Park, G. Park and J. Kim, 2022 International Conference on Robotics and Automation (ICRA), pp. 11409–11415.
19 H. Singh, D. Popov, I. Gaponov and J.-H. Ryu, *Mechanism and Machine Theory*, 2016, **100**, 205–221.
20 C. Xu, T. Liu, S. Dong, Y. Wang and X. Wang, Actuators, 2024, p. 477.
21 Z. Shang and Z. Wang, *Smart materials and structures*, 2012, **21**, 115004.
22 G. C. Tsiatas, I. N. Tsiptsis and A. G. Siokas, International Conference on Vibration Problems, 2019, pp. 857–867.
23 D. Popov, I. Gaponov and J.-H. Ryu, 2012 IEEE/RSJ International Conference on Intelligent Robots and Systems, pp. 1245–1250.
24 H. Singh, D. Popov, I. Gaponov and J.-H. Ryu, 2015 IEEE International Conference on Robotics and Automation (ICRA), 2015, pp. 238–243.